\documentclass[prl,twocolumn,superscriptaddress]{revtex4}
\usepackage{amsmath}
\usepackage{amsfonts}
\usepackage{mathrsfs}
\usepackage{graphicx}
\usepackage{times}
\usepackage{color}

\begin{document}

\title{Detecting $\pi$-phase superfluids with $p$-wave symmetry in a quasi-1D optical lattice}
\author{Bo Liu}
\affiliation{Department of Physics and Astronomy, University of
Pittsburgh, Pittsburgh, PA 15260, USA}
\affiliation{Wilczek Quantum
Center, Zhejiang University of Technology, Hangzhou 310023, China}
\author{Xiaopeng Li}
\affiliation{Condensed Matter Theory Center and Joint Quantum Institute, University of
Maryland, College Park, MD 20742, USA}
\author{Randall G. Hulet}
\affiliation{Department of Physics and Astronomy and Rice Quantum Institute, Rice
University, Houston, TX 77005, USA}
\author{W. Vincent Liu}
\affiliation{Department of Physics and Astronomy, University of
Pittsburgh, Pittsburgh, PA 15260, USA}
\affiliation{Wilczek Quantum
Center, Zhejiang University of Technology, Hangzhou 310023, China}

\begin{abstract}
{We propose an experimental protocol to study  $p$-wave
superfluidity in a spin-polarized cold Fermi gas tuned by an $s$-wave
Feshbach resonance.   A crucial ingredient is to add a quasi-1D
optical lattice and tune the fillings of two spins to the $s$ and
$p$ band, respectively.  The pairing order parameter is confirmed to
inherit $p$-wave symmetry in its center-of-mass motion.  We find
that it can further develop into a state of  unexpected $\pi$-phase
modulation in a broad   parameter regime.    Measurable quantities
are calculated, including time-of-flight distributions,
radio-frequency spectra, and \textit{in situ} phase-contrast imaging
in an external trap. The $\pi$-phase $p$-wave superfluid is
reminiscent of  the $\pi$-state in superconductor-ferromagnet
heterostructures but differs in symmetry  and origin.   If observed,
it would represent another example of $p$-wave pairing, first discovered in He-3 liquids.}
\end{abstract}

\maketitle

{Coexistence of singlet $s$-wave superconductivity with
ferromagnetism is a long-standing issue in condensed matter
physics~\cite{2005_Buzdin_RevModPhys}. One of the most interesting
phenomena is the so-called $\pi$-phase achieved in artificially
fabricated heterostructures of ferromagnetic and superconducting
layers~\cite{2001_Ryazanow_PhysRevLett,2002_Knotos_PhysRevLett,2003_Sellier_PhysRevB,2006_Oboznow_PhysRevLett},
where the relative phase of the superconducting order parameter
between {neighboring superconducting layers} is $\pi$. The
$\pi$-state offers new ways for studying the interplay between
superconductivity and magnetism and has potential application {for
quantum computing in building up superconducting qubits through the
$\pi$ phase shift~\cite{1999_Mooij_science,1999_Ioffe_Nature}.
Different settings for its realization have been discussed, such as
in high $T_c$
superconductors~\cite{1999_Bernhard_PhysRevB,1999_McLaughlin_PhysRevB,2000_Chmaissem_PhysRevB}
and in spin-dependent optical
lattices~\cite{2010_zapata_PhysRevLett}.} }

{{ {In this letter, first we show that an unconventional $p$-wave
$\pi$-phase superfluid state emerges in the experimental system of a
Fermi gas in a quasi-one dimensional optical
lattice~\cite{2010_Hulet_nature}. Then we propose experimental
protocols for observing this novel state by tuning the spin
polarization.  This is reminiscent of the $\pi$-state in
superconductor-ferromagnet
heterostructures~\cite{2005_Buzdin_RevModPhys}.}} {However, the
$\pi$ phase shift of the superfluid gap here arises from a different
mechanism---the relative inversion of the single particle band
structures {($s$- and $p$-orbital bands)} of  {the two spin
components involved} in the pairing.} As a result of this novel
pairing mechanism, such a $\pi$-phase superfluid state has a
distinctive feature---a center-of-mass (COM) $p$-wave symmetry, {
which distinguishes it from other $\pi$-states in previous
studies~\cite{2005_Buzdin_RevModPhys,2010_zapata_PhysRevLett}. We
map out the phase diagram as a} function of controllable
experimental parameters---atom density and spin polarization. There
is a large window for the predicted COM $p$-wave $\pi$-phase
superfluids in the phase diagram at {low density} and it occurs at
higher critical temperature in relative scales, {enhancing its
potential for experimental realization.} {Striking signatures are
calculated for experimental detection: (1) the shape of the density
distribution in time-of-flight shows dramatic changes resulting from
the pairing between different parity orbitals (i.e., $s$ and $p$
orbitals); (2) {distinctive} features are found in the density of
states (DOS) such as the existence of the finite gap and midgap
peak, which can be detected via radio frequency (rf) spectroscopy;
(3) {in the presence of a shallow trap background, } the density
(including atom number density and pair density) distributions in
real space exhibit shell structures composed of the predicted
$p$-wave superfluids.}} The orbital degrees of freedom play an
essential role here; recently the research of higher orbital bands
in optical lattices has evolved
rapidly~\cite{2011_Vincent_naturephy}. For $p$-band fermions with
attractive interaction, the chiral center-of-mass $p$-wave
superfluidity in 2D~\cite{2014_Bo_arxiv} and superfluids similar to
the Fulde-Ferrell-Larkin-Ovchinnikov~\cite{2011_Zicai_PRA} were
found in theoretical studies. {As we shall show with the model
below, the pairing composed of different parity orbital fermions
will lead to unexpected COM $p$-wave $\pi$-phase superfluids.}

\begin{figure}[t]
\begin{center}
\includegraphics[scale=0.25]{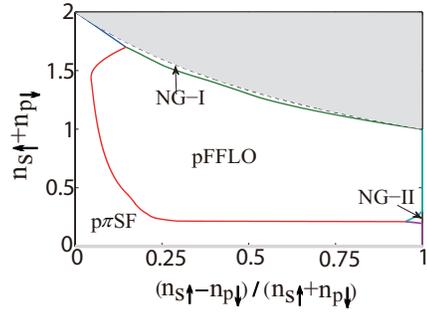}
\end{center}
\caption{Zero temperature phase diagram as a function of lattice
filling and polarization when $t_p/t_s=8$, $t^{\prime }_p/t_s=0.05$,
$t^{\prime }_s/t_s=0.05$ and $U/t_s=-9$. p$\pi$SF and pFFLO stand
for different modulated COM $p$-wave superfluid states with the
center-of-mass momentum of Cooper pairs located at
${\mathbf{Q}}=(\protect\pi/a,0,0) $ and
${\mathbf{Q}}\neq(\protect\pi/a,0,0)$, respectively. NG-I refers to
a normal gas (without pairing) where the $|s\uparrow\rangle$ band is
fully filled while the $|p\downarrow\rangle$ band is partially
filled. NG-II is another kind of normal state where the
$|s\uparrow\rangle$ band is partially filled while the
$|p\downarrow\rangle$ band is nearly empty. The grey area is
forbidden due to the Fermi statistics constraint on the lattice
filling. The grey line stands for the empty state.}
\label{zerophase}
\end{figure}

\textit{Effective model.} Consider a Fermi gas with $s$-wave
attraction composed of two hyperfine states, to be referred to as spin $%
\uparrow $ and $\downarrow $, loaded in a strongly anisotropic 3D
cubic optical lattice. In particular, we consider the lattice
potential {$V_{\rm OL}=\sum\limits_{\alpha =x,y,z}V_{\alpha }\sin^{2}(k_{L}{\mathbf{r}%
_{\alpha }})$ with lattice strengths $V_{z}=V_{y}\gg V_{x}$, where
$k_L$ is the wavevector of the laser fields}. In the deep lattice
limit, the lattice potential at each site can be approximated by a
harmonic oscillator and the lowest two energy levels are $s$ and
$p_x$ orbital states for the lattice potential considered here,
{while other  levels such as $p_y$ and $p_z$ orbitals are well
separated in energy}. In the following, the $p_x$ orbital state is
simply called the $p$ orbital. Due to the strong confinement of the
lattice potential in $y$ and $z$ directions, the system is
dynamically separated into an array of quasi-one dimensional tubes.
Further, let the gas be tuned with spin imbalance by the techniques
developed in recent experiments~\cite{2006_Partridge_science,2006_Martin_Science,2010_Salomon_nature}.
{A key condition proposed here is to tune the imbalance sufficiently
that the spin $\uparrow $ and $\downarrow $ Fermi levels
reside in the $s$ and $p$ orbital bands respectively in order to
hybridize the spin and orbital degrees of freedom.}  In the tight
binding regime, the system is described by a multi-orbital Fermi
Hubbard model
\begin{eqnarray}
\textstyle
H &=&-t_{s}\sum_{\mathbf{r}}C_{s\uparrow}^{\dagger }(\mathbf{r})C_{s\uparrow}(\mathbf{r}+\vec{%
e}_{x})+t_{p}\sum_{\mathbf{r}}C_{p\downarrow}^{\dagger }(\mathbf{r})C_{p\downarrow}(\mathbf{r}+\vec{e}%
_{x}) \notag \\
&-&t^{\prime}_{s}\sum_{\mathbf{r}}[C_{s\uparrow}^{\dagger }(\mathbf{r})C_{s\uparrow}(%
\mathbf{r}+\vec{e}_{y})+C_{s\uparrow}^{\dagger }(\mathbf{r})C_{s\uparrow}(\mathbf{r}+\vec{e}%
_{z})] \notag \\
&-&t^{\prime}_{p}\sum_{\mathbf{r}}[C_{p\downarrow}^{\dagger }(\mathbf{r})C_{p\downarrow}(%
\mathbf{r}+\vec{e}_{y})+C_{p\downarrow}^{\dagger }(\mathbf{r})C_{p\downarrow}(\mathbf{r}+\vec{e}%
_{z})]+h.c.  \notag \\
&-&\mu _{\uparrow}\sum_{\mathbf{r}}C_{s\uparrow}^{\dagger }(\mathbf{r})C_{s\uparrow}(\mathbf{r}%
)-\mu _{\downarrow}\sum_{\mathbf{r}}C_{p\downarrow}^{\dagger }(\mathbf{r})C_{p\downarrow}(\mathbf{r}%
)\notag \\
&+&U\sum_{\mathbf{r}}C_{s\uparrow}^{\dagger }(\mathbf{r})C_{s\uparrow}(\mathbf{r}%
)C_{p\downarrow}^{\dagger }(\mathbf{r})C_{p\downarrow}(\mathbf{r}),
\label{Ham}
\end{eqnarray}
{where} $t_s$ and $t_p$ are the hopping amplitudes along the $x$
direction for
the $s$ and $p$ band fermions, respectively, while $t^{\prime }_s$ and $%
t^{\prime}_p$ are the hopping amplitude along the $y$ and $z$
directions. {All the hopping amplitudes as introduced in
Eq.~\eqref{Ham} are positive and the relative {signs} before them
{are} fixed by the parity symmetry of the $s$ and $p$ orbital wave
functions.} $C_{\nu \sigma}(\mathbf{r})$ is a fermionic annihilation
operator for the spin $\sigma$ component ($\uparrow$ and
$\downarrow$) fermion with the localized $\nu$ ($s$ and $p$) orbital
located at the lattice site ${\mathbf{r}}$, and $\mu_{\sigma}$ is
the corresponding chemical potential. {The onsite interaction (last
term in Eq.~\eqref{Ham}) is of the density-density type and arises
from the interaction between two hyperfine states, which is highly
tunable through the $s$-wave Feshbach resonance in the ultracold
atomic gases.} {Here we assume that the interaction strength is much
smaller than the band gap. Therefore, the $s$-band fully filled spin
down fermions are dynamically inert which are not included in the
Hamiltonian (Eq.~\eqref{Ham}).} In this work, we focus on the case
with attractive interaction where superfluidity is energetically
favorable.

\textit{Phase diagram at zero temperature.} In order to study the
superfluidity in our system, we apply the mean-field approximation
and assume the superfluid pairing is between different parity orbitals,
i.e., between $|s\uparrow\rangle$ and $|p\downarrow\rangle$ states,
in a general form $\Delta (\mathbf{r})=U\langle
C_{p\downarrow}(\mathbf{r)}C_{s\uparrow}(\mathbf{r)}\rangle
=\sum\limits_{m =1}^{M}\Delta _{m}\exp (i\mathbf{Q}_{m}\cdot\mathbf{%
 r})$, where $M$ is an integer. Then, the Hamiltonian in the momentum space
reads
\begin{eqnarray}
 \textstyle
 H_{MF} &=&\sum\limits_{\mathbf{k}}\varepsilon
_{s}(\mathbf{k})C_{s\uparrow}^{\dagger
}(\mathbf{k})C_{s\uparrow}(\mathbf{k})+\sum\limits_{\mathbf{k}}\varepsilon _{p}(%
\mathbf{k})C_{p\downarrow}^{\dagger }(\mathbf{k})C_{p\downarrow}(\mathbf{k}) \notag \\
&+&\sum\limits_{m =1}^{M}\sum\limits_{\mathbf{k}}(\Delta _{m}^{\ast
}C_{p\downarrow}(-\mathbf{k+Q}_{m})C_{s\uparrow}(\mathbf{k})+h.c.) \notag \\
&-&N\sum\limits_{m =1}^{M}\frac{|\Delta _{m}|^{2}}{U}, \label{MHam}
\end{eqnarray}
where $\varepsilon _{s}(\mathbf{k})=-2t_{s}\cos
k_{x}a-2t_{s}^{\prime }(\cos k_{y}a+\cos k_{z}a)-\mu
_{\uparrow},\varepsilon _{p}(\mathbf{k})=2t_{p}\cos
k_{x}a-2t_{p}^{\prime}(\cos k_{y}a+\cos k_{z}a)-\mu
_{\downarrow},\mathbf{k}$ is the lattice momentum in the first
Brillouin zone (BZ), {$N$ is the total number of lattice sites and
$a$ is the lattice constant.

A fully self-consistent mean field calculation for the
space-dependent order parameter is numerically challenging. We
restrict our discussion to {two forms of a variational ansatz}. The
first one is analogous to the Fulde-Ferrell state, where it assumes
that
{$M=1$} and the order parameter reads $\Delta (\mathbf{r})=\Delta \exp (i%
\mathbf{Q\cdot r})$. The absolute value of superfluid gap is a
constant, but the phase alternates from site to site. The second one
is chosen by {$M=2$}, $\Delta _{1}=\Delta _{2}=\frac{\Delta }{2}$ and $\mathbf{%
Q}_{1}=-\mathbf{Q}_{2}$, namely $\Delta (\mathbf{r})=\Delta \cos (\mathbf{%
Q\cdot r})$, which is an analogue of the Larkin-Ovchinnikov state.
This variational approach adopted here was previously justified by
density-matrix-renormalization-group
methods~\cite{2010_Zixu_PhysRevA}.} Here we choose
$\mathbf{Q}$ pointing along the $x$ direction, say $\mathbf{Q}=\mathrm{Q}%
(1,0,0)$ to fully gap the Fermi surface of this quasi-one
dimensional system. {For  FF like states,}  the mean-field
Hamiltonian in Eq.~\eqref{MHam} can be diagonalized through the
Bogoliubov transformation. The grand canonical potential of the
system reads
\begin{eqnarray}
\textstyle
 \Omega &=&\sum\limits_{\mathbf{k}}[\sum\limits_{\gamma
=\pm }-\frac{1}{\beta
}\ln (1+\exp (-\beta \zeta _{\gamma }(\mathbf{k}%
)))+\varepsilon _{p}(\mathbf{k})]-N\frac{|\Delta |^{2}}{U},\notag \\
\label{Freeenergy}
\end{eqnarray}
 where $\zeta _{\pm }(\mathbf{k})=\frac{1}{2}[\varepsilon _{s}(\mathbf{%
k})-\varepsilon _{p}(\mathbf{-k+Q})\pm \sqrt{\lbrack \varepsilon _{s}(\mathbf{%
k})+\varepsilon _{p}(\mathbf{-k+Q})]^{2}+4|\Delta |^{2}}]$ is the
quasi-particle dispersion. While for the case of {LO-like} states,
the Hamiltonian can not be diagonalized analytically. {However, the
pairing terms {link}  the one-particle states with momenta lying a
single line in the BZ. Therefore, we can solve the resulting
eigenproblem numerically for relatively large {systems}.}

From our numerics, we find that the free energy of the analogous LO
states is always {lower} than that of the FF like phases, except at
${\mathbf{Q}}=(\pi/a,0,0)$, where the FF and LO-like ansatz are
equivalent. So the ground state of the system is a COM $p$-wave
superfluid state with modulated pairing order parameter
$\propto\cos(\mathbf{Q\cdot r})$, which breaks the translational
symmetry spontaneously. Qualitatively, that is because the $\pm
{\mathbf{Q}}$ pairing opens gaps on both sides of the Fermi surface,
taking advantage of the available phase space for pairing, while the
FF states only open a gap on one side. {{Since the dispersion of the
$p$ band is inverted with respect to {that of} the $s$ band, the
pairing occurs between fermions with center-of-mass momentum
$Q\simeq k_{F \uparrow} + k_{F \downarrow}$, where $k_{F \uparrow}$
and $k_{F \downarrow}$ are the two relevant Fermi momenta. When the
occupation numbers of $|s\uparrow\rangle$ and $|p\downarrow\rangle$
states are equal, the $\pi$-phase superfuild state with
${\mathbf{Q}}=(\pi/a,0,0)$ is the ground state of the system. {In
real space, the pairing order parameter is a function of staggered
signs along the $x$ direction and obeys $\Delta({\mathbf
r})=-\Delta({\mathbf r}+a{\mathbf e_x})$ when combined with the
periodicity $\Delta({\mathbf r})=\Delta({\mathbf r}+2a{\mathbf
e_x})$.} The $\pi$ phase shift of the superfluid gap here arises
from the relatively inverted single particle band structures
directly, unlike in the conventional FFLO
state~\cite{2011_Leo_PhysRevA}. The predicted $\pi$-phase superfluid
state is found to be quite robust. Even when the occupation number
difference between $|s\uparrow\rangle$ and $|p\downarrow\rangle$
states is finite, the $\pi$-phase superfuild state is still the
ground state. Particularly in the low density region $n_s + n_p \ll
1$, there is a large window for this $\pi$-state.}

{{When the polarization
$p=\frac{n_{s\uparrow}-n_{p\downarrow}}{n_{s\uparrow}+n_{p\downarrow}}$
is sufficiently large,} the center-of-mass momentum will become
incommensurate with the underlying lattice and this incommensurate
COM $p$-wave state will be referred to as pFFLO. {As shown in
Fig.~\ref{zerophase}, a phase diagram as a function of atom density
and polarization has been obtained. When the polarization {is} below
a critical value, the system favors the $\pi$-phase superfluids.
{Beyond this value, a first order phase transition occurs from the
$\pi$-phase to pFFLO superfluid states.} {It is worth to note that a
large regime of parameters is found to exist in the phase diagram,
making the experimental realization of the predicted new $p$-wave
pairing phases simpler.}}}

Here we would like to stress two distinctions of the predicted
$p$-wave superfluid states with different types of modulated pairing
from the conventional $s$-band FFLO~\cite{2008_Batrouni_PhysRevLett}.
One is that the Cooper pairs
here are composed of different parity orbital fermions (i.e., $s$
and $p$ orbitals), which lead to the pairing order parameter with
$p$-wave symmetry in the COM motion. {The other is that the pairing
order is modulated with the wavevector $Q\simeq k_{F \uparrow} +
k_{F \downarrow}$, rather than determined} by the Fermi surface
mismatch.}

\begin{figure}[t]
\includegraphics[scale=0.25]{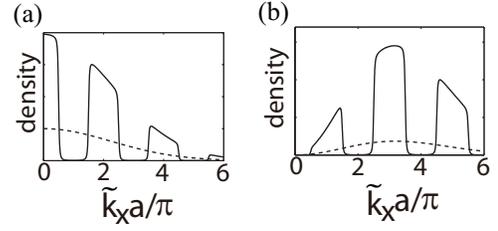}
\caption{Prediction of spin-resolved density distribution in
time-of-flight, in (a) for $|s\uparrow\rangle$ fermions, while in
(b) for $|p\downarrow\rangle$ fermions. The solid line shows the
density defined in the main text along the $\tilde{k}_x$-axis. The
dash lines show the intensities of the Wannier orbital functions
$\propto |\phi_{\nu}({\tilde{k}_x})|^2$ for comparison. {Other
parameters are $n_{s\uparrow}=0.5$, $n_{p\downarrow}=0.5$,
$t_p/t_s=8$, $t^{\prime}_p/t_s=0.05$, $t^{\prime}_s/t_s=0.05$ and
$U/t_s=-9$.}} \label{tof}
\end{figure}

\textit{Experimental signatures.} {The most distinctive feature of
the predicted $\pi$-phase and pFFLO superfluids is that the pairing
order parameters are spatially modulated and have $p$-wave
symmetry in the COM motion. This leads to several characteristic
experimental signatures.} (A) the single particle momentum
distributions exhibit unique properties in the following two
aspects. The {\it first} one is the shape of the density
distribution in time-of-flight. {We calculate the
spin-resolved (or equivalently orbital-resolved) density
distribution in the time-of-flight measurement assuming ballistic
expansion as $\langle\tilde{n}_{\nu\sigma}(x)\rangle_t
=\left(\frac{m}{\hbar t}\right)^2 \sum_{\tilde{k}_y,\tilde{k}_z}
\phi^*_{\nu} ({\tilde{\mathbf k}}) \phi_{\nu} ({\tilde{\mathbf k}})
\langle C^\dagger_{\nu\sigma} ({{\mathbf k}}) C_{\nu\sigma}
({{\mathbf k}})\rangle$, where ${\tilde{\mathbf k}}=m {\mathbf
r}/(\hbar t)$, $\phi_{\nu}({\tilde{\mathbf k}})$ is the Fourier
transform of the $\nu$-orbital band Wannier function
$\phi_{\nu}({\mathbf r})$ and $\mathbf {k} = {\tilde{\mathbf k}}$
mod $\mathbf G$ is the momentum in the first BZ corresponding to
${\tilde{\mathbf k}}$ ($\mathbf G$ is the primitive reciprocal
lattice vector).} As shown in Fig.~\ref{tof}, the highest peak for
$p$ band fermions is shifted from zero momentum resulting from
the non-trivial profile of the $p$-wave Wannier function superposed on
the density distributions, distinguished from the $s$ orbital
fermions. {The {\it second} aspect is a mirror-translational
symmetry of the axial density distributions of $|s\uparrow\rangle$
and $|p\downarrow\rangle$ fermions for the $\pi$-phase superfluid
state. Following the standard
analysis~\cite{2006_Silva_PhysRevLett,2010_Hulet_nature}, we define
the  axial density distribution in the momentum space as
$n^{a}_{\nu\sigma}(k_{x})=\frac{1}{(2\pi )^{2}}\int
dk_{y}dk_{z}\langle C^\dagger_{\nu\sigma}({\mathbf k})
C_{\nu\sigma}({\mathbf k})\rangle$ for $|s\uparrow\rangle$ and
$|p\downarrow\rangle$ fermions, respectively. {From
Eq.~\eqref{Freeenergy}, the atom density distributions of
$|s\uparrow\rangle$ and $|p\downarrow\rangle$ fermions can be
calculated through $n_{\nu\sigma}({\mathbf
k})=\frac{\partial\Omega}{\partial\mu_{\sigma}}$. In the zero
temperature, we find that $n_{s\uparrow}({\mathbf k})=n_{p
\downarrow}({\mathbf {Q-k}})=\frac{-[\varepsilon
_{s}(\mathbf{k})+\varepsilon
_{p}(\mathbf{-k+Q})]+\sqrt{\mathbb{E}(\mathbf{k})}}{2\sqrt{\mathbb{E}(\mathbf{k})}}$
where $\mathbb{E}(\mathbf{k})=[\varepsilon
_{s}(\mathbf{k})+\varepsilon _{p}(\mathbf{-k+Q})]^{2}+4|\Delta
|^{2}$ and ${\mathbf Q}=(\pi/a,0,0)$. Therefore, the axial density
distributions defined above satisfy the relation
$n^{a}_{s\uparrow}(k_x)=n^{a}_{p\downarrow}(\pi/a-k_x)$.} It is also
confirmed {in} our numerics as shown in Fig.~\ref{DOS}(c). These
signatures can be detected through polarization phase contrast
imaging~\cite{2010_Hulet_nature}.}

\begin{figure}[t]
\begin{center}
\includegraphics[scale=0.23]{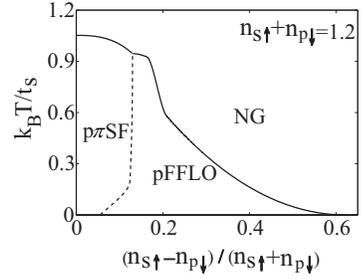}
\end{center}
\caption{Finite temperature phase diagram as a function of
temperature and polarization at a certain lattice filling when
$t_p/t_s=8$, $t^{\prime }_p/t_s=0.05$, $t^{\prime }_s/t_s=0.05$ and
$U/t_s=-9$.} \label{finitephase}
\end{figure}

{(B) the COM $p$-wave superfluid state here has the spatially
nonuniform pairing order parameter, making it different from
conventional $p$-wave superfluids.} {This leads to crucial
difference in Bogoliubov quasi-particle spectra.} {A finite energy
gap is shown in the spin-resolved (or equivalently the
orbital-resolved) density of states (DOS) for the $\pi$-phase
superfluids (Fig.~\ref{DOS}a). Such spin-resolved DOS is calculated
as $\rho _{\nu \sigma}(E)=\frac{1}{2}\sum_{n}[|u_{n}^{\nu
\sigma}|^{2}\delta (E-\zeta _{n})+|v_{n}^{\nu \sigma}|^{2}\delta
(E+\zeta _{n})]$, where $(u_{n}^{\nu \sigma},v_{n}^{\nu
\sigma})^{T}$ is the eigenvector corresponding to the eigenenergy
$\zeta _{n}$ of the mean-field Hamiltonian Eq.~\eqref{MHam} and the
summation runs over all the eigenenergy. This finite gap in the DOS
gives direct evidence of superfluidity. {For the pFFLO
state, there is a midgap peak in the DOS, shown in Fig.~\ref{DOS}(b).}
This midgap peak results from the spatially modulated pairing order
parameter which {signifies the emergence of} Andreev bound
states~\cite{2008_Toma_PRL}. {The energy gap and midgap peak are
found in the spin-resolved DOS for both $|s\uparrow\rangle$ and
$|p\downarrow\rangle$ fermions.} For example, the DOS of
$|s\uparrow\rangle$ fermions is shown in Fig.~\ref{DOS}. Such
spin-resolved DOS signatures can be detected via
{radio frequency (rf)} spectroscopy~\cite{2003_Ketterle_Science,2003_Jin_PRLrf,2007_Ketterle_PRL},
giving an experimental plausible probe of the predicted $p$-wave
superfluids.

\begin{figure}[t]
\begin{center}
\includegraphics[scale=0.28]{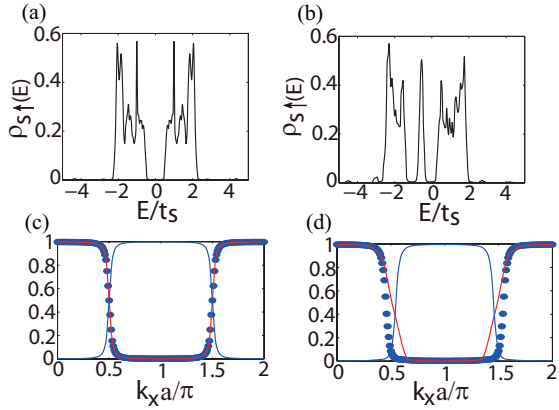}
\end{center}
\caption{Top row plots density of states (DOS)
$\rho_{s\uparrow}(E)$, (a) showing the finite energy gap for  the
$\pi$-phase superfluid state and (b) showing the midgap peak for
pFFLO  resulting from Andreev bound states. Bottom row plots the
axial density distributions of $|s\uparrow\rangle$ and
$|p\downarrow\rangle$ fermions in momentum space for each of
$\pi$-phase (c) and pFFLO (d). {The red and blue solid lines show
$n^{a}_{s\uparrow}(k_x)$ and $n^{a}_{p\downarrow}(k_x)$
respectively, while the blue dots show
$n^{a}_{p\downarrow}(Q-k_x)$.} See main text for the definition of
$\rho_{s\uparrow}(E)$ and $n^{a}_{\nu\sigma}(\mathbf{k})$. {For the
pFFLO state in (b), since there is a large polarization which can be
considered as an effective external magnetic field, it leads to a
shift of the density of states. Therefore, the midgap peak in (b) is
not at $E=0$.} {In (a) and (c), we choose $n_{s\uparrow}=0.5$ and
$n_{p\downarrow}=0.5$, while in (b) and (d), $n_{s\uparrow}=0.52$
and $n_{p\downarrow}=0.45$. Other parameters are the same as in
Fig.~\ref{zerophase}.}} \label{DOS}
\end{figure}

{(C) the} predicted COM $p$-wave superfluids require neither
spin-orbital coupling nor an induced {second order} effective
$p$-wave interaction. It arises directly from a purely $s$-wave
two-body attraction. This leads to a significantly improved
transition temperature, which is confirmed {by our direct
calculation} of finite temperature phase transitions for the model
Hamiltonian in Eq.~\eqref{Ham}---see Fig.~\ref{finitephase} for the
phase diagram. The transition temperature of the predicted COM
$p$-wave superfluids is {indeed  significantly improved as compared
to} other conventional relative $p$-wave
superfluids~\cite{2007_Leo_AnnalsPhy,2005_Iskin_PhysRevB}.} {For
instance, suppose a quasi-1D optical lattice has the lattice depth
$V_x=5E_r, V_y=V_z=18E_r$ and the wavelength of the lattice beams is
$1064$nm, {the mean-field superfluid transition temperature} can
reach around $60$nK when the s-wave scattering length between $^6$Li
atoms is $a_s \simeq 326 a_0$~\cite{2015_Hart_nature}, where $a_0$
denotes the Bohr radius. Further increasing $a_s$, to around $600
a_0$, the transition temperature can reach around $0.2\mu$K, or even
higher.}

\begin{figure}[t]
\begin{center}
\includegraphics[scale=0.26]{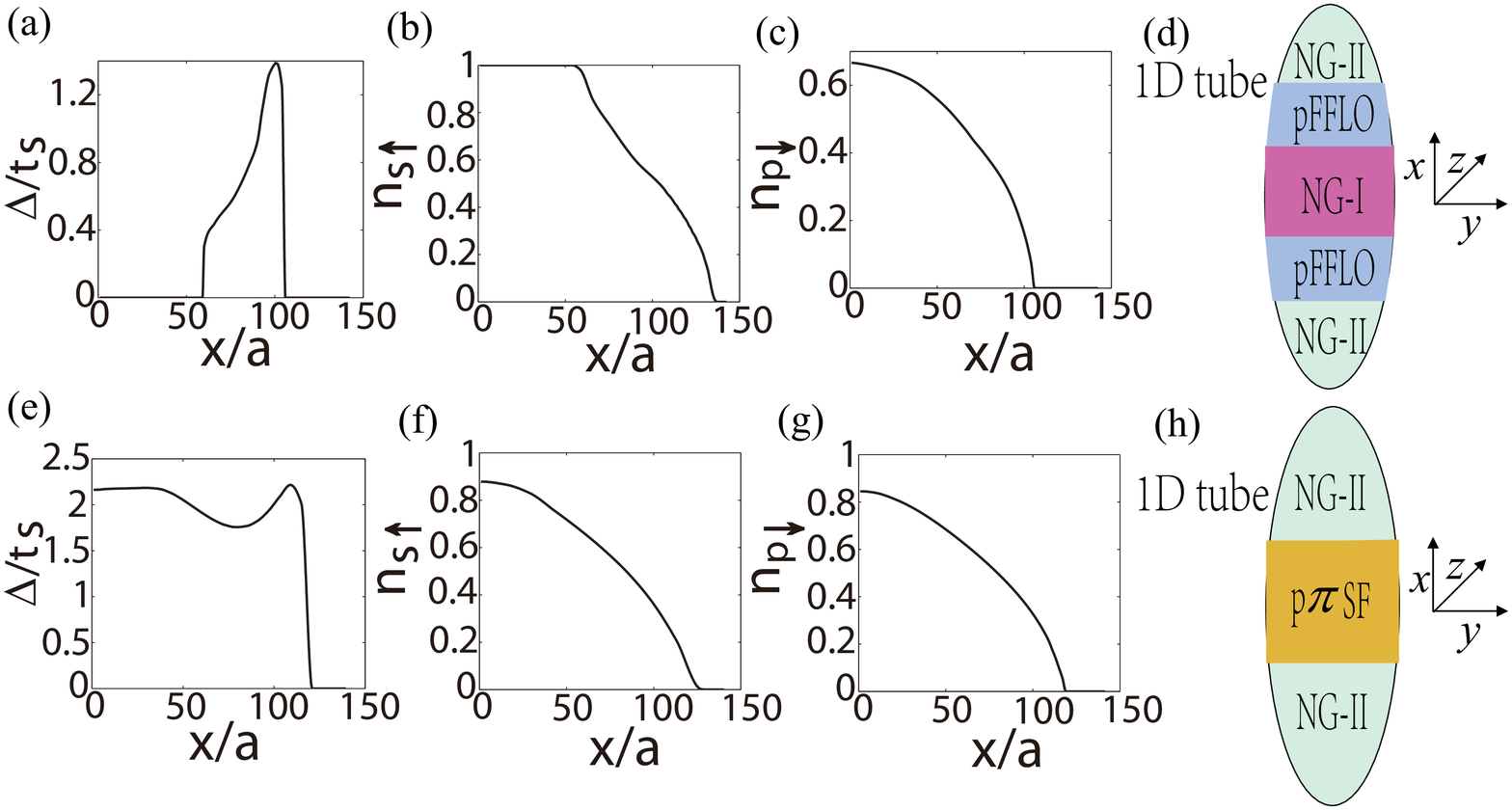}
\end{center}
\caption{Shell structures with a background trapping potential. The
superfluid gap $\Delta$ and density profile $n_{s\uparrow}$ and
$n_{p\downarrow}$ of $|s\uparrow\rangle$ and $|p\downarrow\rangle$
fermions are shown as a function of the coordinate $x$, {when
$t_p/t_s=8$, $t^{\prime }_p/t_s=0.05$, $t^{\prime
}_s/t_s=0.05$ and $U/t_s=-9$.} The polarization $P=\frac{N_{s\uparrow}-N_{p\downarrow}}{%
N_{s\uparrow}+N_{p\downarrow}}$ is fixed at $0.3$ and $0.03$ for the
first and second row, respectively. The frequency of the harmonic
trap is chosen to be $120$ Hz.} \label{trap}
\end{figure}

\textit{Shell structure in a trap.} {In the following, we will
discuss the effect of a harmonic trapping potential superposed on
the optical lattices. Assuming that the harmonic trapping potential
is sufficiently shallow compared with the lattice depth, it is
{natural} to apply the local density approximation (LDA) and let the
chemical potential vary as a function of the position. Here we
consider {the trapping potential in the $x$ direction.}
Fig.~\ref{trap} shows various shell structures by our calculation.
When the polarization is small, the region in the center of trap is
{the $\pi$-phase superfluid}, surrounded by a normal gas shell.
While increasing the polarization, the region in the center of trap
is no longer a superfluid, but a normal gas, surrounded by a pFFLO
superfluid shell, which in turn is surrounded by another shell of a
normal gas. The density profile of $|s\uparrow\rangle$ and
$|p\downarrow\rangle$ fermions along the $x$ direction are also
demonstrated in Fig.~\ref{trap}, which can be detected through
\textit{in situ} phase-contrast imaging, providing an experimental
plausible probe in a trapped atomic Fermi
gas~\cite{2010_Hulet_nature}.}

{Let us take a specific experimental system, $^6$Li atoms, as an
example. This Fermi gas is typically prepared with a total atom
number around $10^5$ in a
trap~\cite{2010_Hulet_nature,2006_Partridge_science}.  In the recent
experiment, it  is loaded in an optical lattice and cooled down to
the antiferromagnetic (N\'{e}el) transition
scale~\cite{2015_Hart_nature}.  At the center of the trap, the
density of one particle per site is maintained.  One can take
advantage of this experimental development to achieve the
$\pi$-phase $p$-wave state.   One apparent method is to have a
higher density, say, through tightening of the trap potential or
filling more atoms into the lattice
initially~\cite{2005_Michael_PhysRevLett,2008_Schneider_science,2008_Jordens_nature}.
{Taking the peak density of atoms to be around $2\times 10^{13}
cm^{-3}$, the central filling of the lattice system is three
particles per site when the lattice constant $a=532$nm.} Another
method is through tuning the polarization parameter of the two
hyperfine-spin components, so to populate one and only one of the
two components of atoms into the $p$-orbital band. The two methods
may be combined to optimize the experiment.}

\textit{Conclusion.} {We propose that the pairing between different
parity orbital fermions can lead to a $p$-wave $\pi$-phase
superfluid state. {The origin of the $\pi$ phase shift of the pairing
order is distinct from the previous studies of $\pi$-states.}  {We
show that the predicted $\pi$-phase here occurs in a broad range in
the phase diagram especially in the low density region. Increasing
polarization, we find a phase transition from the $\pi$-phase state
to an incommensurate COM $p$-wave superfluid.} Signatures of the
predicted $p$-wave superfluid states are calculated for
time-of-flight and spectroscopic measurements in the regime of atom
density and spin polarization deemed experimentally accessible. The
transition temperature and various shell structures with a
background trapping potential {are obtained for future } experiments
to explore these new forms of $p$-wave superfluid states.}

\textit{Acknowledgements.} This work is supported by AFOSR
(FA9550-12-1-0079), ARO (W911NF-11-1-0230), Overseas Collaboration
Program of NSF of China No. 11429402 sponsored by Peking University,
the Charles E. Kaufman Foundation, and The Pittsburgh Foundation (B.
L. and W. V. L.). X. L. is supported by LPS-MPO-CMTC, JQI-NSF-PFC
and ARO-Atomtronics-MURI. R. G. H. acknowledges support from ARO
Grant No. W911NF-13-1-0018 with funds from the DARPA OLE program,
NSF, ONR, the Welch Foundation (Grant No. C-1133), and ARO-MURI
Grant No. W911NF-14-1-0003.

\bibliographystyle{apsrev}
\bibliography{comsppair}

\end{document}